\documentclass[journal]{IEEEtran}
\usepackage{amssymb}
\usepackage{graphicx}
\usepackage{epstopdf}
\usepackage{gensymb}
\usepackage{color}
\usepackage{amsmath}
\usepackage{bm}
\usepackage{array}
\usepackage{booktabs}
\usepackage{multirow}
\usepackage{threeparttable} 
\usepackage{mathtools}

\usepackage{amsthm}

\ifCLASSINFOpdf
\else
\fi
\begin{document}
%
\title{PUF-FSM: A Controlled Strong PUF}
%
%
%

\author{
	Yansong Gao and Damith C.~Ranasinghe
 
\thanks{Y.~Gao and D.~C Ranasinghe are with the Auto-ID Labs, School of Computer Science, The University of Adelaide, SA 5005, Australia. e-mail: \{yansong.gao, damith.ranasinghe\}@adelaide.edu.au.}
}

\markboth{XXXX}%
{Shell \MakeLowercase{\textit{et al.}}: Bare Demo of IEEEtran.cls for Journals}

\maketitle

\begin{abstract}
Physical unclonable functions (PUF), as hardware security primitives, exploit manufacturing randomness to extract instance-specific challenge (input) response (output) pairs (CRPs). Since its emergence, the community started pursuing a strong PUF primitive that is with large CRP space and resilient to modeling building attacks. A practical realization of a strong PUF is still challenging to date. 
This paper presents the PUF finite state machine (PUF-FSM) that is served as a practical {\it controlled} strong PUF. Previous controlled PUF designs have the difficulties of stabilizing the noisy PUF responses where the error correction logic is required. In addition, the computed helper data to assist error correcting, however, leaks information, which poses the controlled PUF under the threatens of fault attacks or reliability-based attacks. The PUF-FSM eschews the error correction logic and the computation, storage and loading of the helper data on-chip by only employing error-free responses judiciously determined on demand in the absence of an Arbiter PUF with a large CRP space. In addition, the access to the PUF-FSM is controlled by the trusted entity. Control in means of i) restricting challenges presented to the PUF and ii) further preventing repeated response evaluations to gain unreliability side-channel information are foundations of defensing the most powerful modeling attacks. The PUF-FSM goes beyond authentications/identifications  to such as key generations and advanced cryptographic applications built upon a shared key.
\end{abstract}

\begin{IEEEkeywords}
Physical uncloanble function, APUF, error-free responses, statistical model, modeling attacks, fault attacks.
\end{IEEEkeywords}

\IEEEpeerreviewmaketitle

\section{Introduction}\label{Introduction}
Physical unclonable function (PUF), as a hardware security primitive, exploits manufacturing variations to extract secrets on demand~\cite{suh2007physical,herder2014physical}. PUFs are increasingly adopted to provide security for pervasive and ubiquitous distributed resource-constraint smart Internet of Thing (IoT) devices as an alternative to storing a digital secret in the non-volatile memory (NVM). In fact, digital keys in the NVM is vulnerable to various attacks, especially, when there is no dedicated room in cost sensitive IoT devices to implement expensive protection mechanisms. By using a PUF, there is no digital secret---must be securely stored---involved, the hardware itself is the secret key that is originated from true randomness, therefore, the secret cannot be duplicated and holds higher resistance to attacks, especially invasive attacks~\cite{gassend2008controlled}.

Since the first silicon PUF, Arbiter PUF (APUF), being coined in 2002~\cite{gassend2002silicon}, the PUF community has never stopped pursuing on the so-called strong PUFs that not only have a large challenge response pair (CRP) space but also resilient to modeling attacks. Applications of the strong PUF range from elementary identifications and authentications to key generations and more advanced cryptographic protocols such as key exchange and oblivious transfer~\cite{ruhrmair2013pufs}. Though there does exist strong PUFs such as the Optical PUF~\cite{pappu2002physical} and the SHIC PUF~\cite{ruhrmair2011applications}, a practical and lightweight strong PUF realization seamlessly compatible with current CMOS technology turns out to be challenging~\cite{vijayakumar2016machine} in front of modeling attacks such as logistic regression (LR) and recently revealed more powerful Covariance Matrix Adaptation Evolution Strategy (CMA-ES) attacks, which have broken previously deemed practical strong PUFs including XOR-APUF, Feedforward APUF, Lightweight Secure PUF~\cite{ruhrmair2013puf,ruhrmair2010modeling,majzoobi2008testing} and even Slender PUF~\cite{becker2015gap,becker2015pitfalls}.

Yu {\it et al.}~\cite{yulockdown} recently presented a practical strong PUF through upper-bounding the available number of CRPs by an adversary. In this work, gaining new CRPs materials has to be implicitly authorized by the trusted entity, the concept of limiting access to CRPs is alike to controlled PUFs~\cite{gassend2008controlled}, detailed in Section~\ref{Sec:ControllPUF}. Yu {\it et al.}~\cite{yulockdown} further introduce a PUF device side nonce to prevent fault attacks or noise side-channel information based attacks~\cite{becker2015gap,becker2015pitfalls}. 

We continue the efforts into pursuing a practical and lightweight strong PUF coined as the PUF-FSM. For authentication, the PUF-FSM gets around one major limitation of~\cite{yulockdown} in terms of available secure authentication times (rounds). Beyond authentication, it enables key generation, key exchange and more advanced cryptographic applications with {\it no reliance on on-chip ECC and the associated helper data}. Eventually, the PUF-FSM is a {\it practical} controlled PUF realization. Contributions of our work are fourfold:
\begin{itemize}
	\item We present a {\it practical and lightweight strong} PUF realization termed as PUF-FSM, also a controlled strong PUF, enabling a wide spread of applications.
	\item We, for the first time, eschew the ECC and helper data to build a controlled PUF. We only employ large number of available error-free responses in absence of the APUFs.
	\item We post-process the responses to prevent traditional machine learning attacks such as LR that usually requires direct relationship between challenge and response. 
	\item We prevent noise side-channel information based attacks (fault attacks) such as the CMA-ES attacks by using device side nonce inherited from~\cite{yulockdown} to disable observing repeated evaluated responses or outputs when the same challenge is maliciously applied. 
\end{itemize}
Section~\ref{Sec:RelWork} introduces related work, especially, judiciously selection of error-free responses from a statistical APUF model. Section~\ref{Sec:PUF-FSM} details the PUF-FSM design and analyzes its security; Wide spread of applications of the PUF-FSM are presented in Section~\ref{Sec:Discussion}; Section~\ref{Sec:Conclusion} concludes this paper.

\section{Related Work}\label{Sec:RelWork}
\subsection{APUF Model for Error-Free Response Generation}
\subsubsection{Modeling APUF}
\begin{figure} [t]
\centering
\includegraphics[trim=0 0 0 0,clip,width=0.4\textwidth]{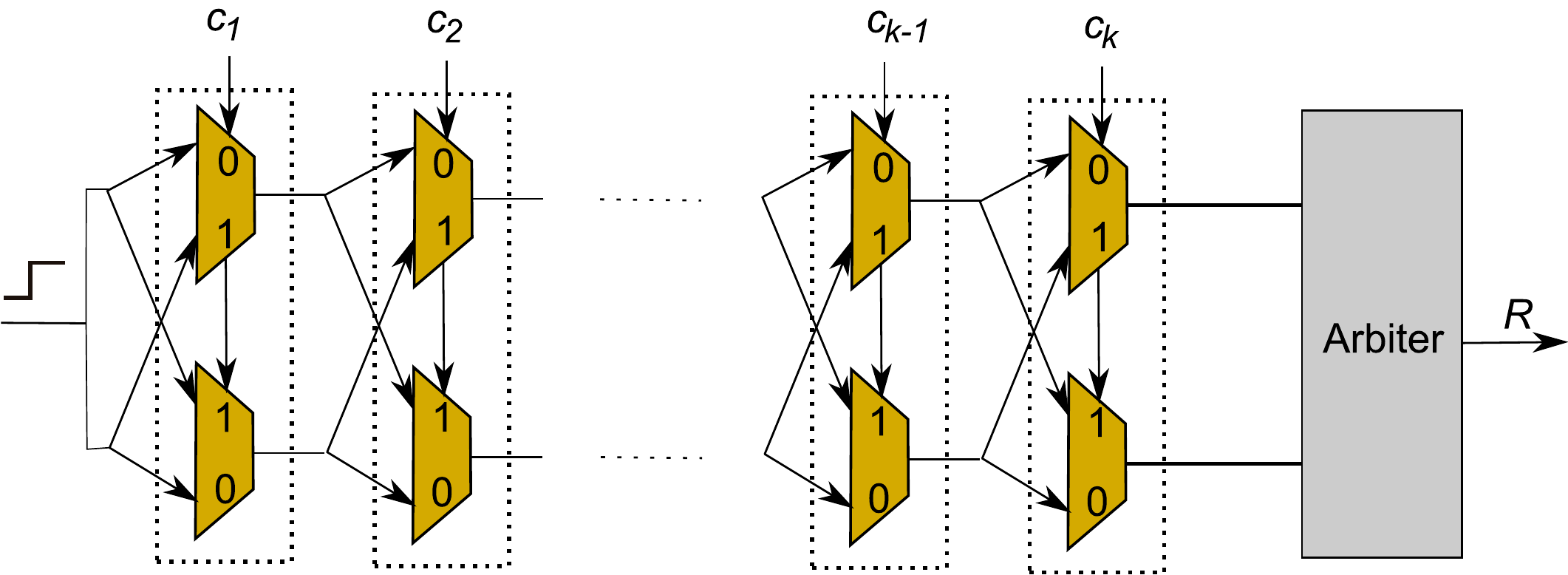}
\caption{An arbiter PUF (APUF) circuit.}
\label{APUF}
\end{figure}

The APUF consists of $k$ stages of two 2-input multiplexers as shown in Fig.~\ref{APUF}, or any other units forming two signal paths. To generate a response bit, a signal is applied to the first stage input, while the challenge $ \bf C $ determines the signal path to the next stage. The input signal will race through each multiplexer path (top and bottom paths) in parallel with each other. At the end of the APUF architecture, an arbiter, e.g., a latch, determines whether the top or bottom signal arrives first and hence results in a logic `0' or `1' accordingly.

It has been shown that an APUF can be modelled via a linear additive model because a response bit is generated by comparing the summation of each time delay segment in each stage (two 2-input multiplexers) depending on the challenge $\bf C$, where $\bf C$ is made up of ($ c_1|| c_2||  ...|| c_k $)~\cite{ruhrmair2013puf,lim2004extracting,ruhrmair2010modeling}. The notations in this section following~\cite{ruhrmair2013puf,ruhrmair2010modeling}. The final delay difference $ t_{\rm dif} $ between these two paths is expressed as:
\begin{equation}
 t_{\rm dif} = {{\boldsymbol \omega}}^T {\bf {\Phi}} ,
\end{equation}

where $ {\boldsymbol \omega} $ and $\bf{ {\Phi}} $ are the delay determined vector and the parity vector, respectively, of dimension $k+1$  as a function of $\bf C$. We denote ${\sigma}_i^{1/0} $ as the delay in stage $i$ for the crossed ($ c_i=1 $) and uncrossed ($ c_i=0$) signal path through the multiplexers, respectively. Hence ${\sigma}_i^{1} $ is the delay of stage $i$ when $ c_i=1 $, while ${\sigma}_i^{0}$ is the delay of stage $i$ when $ c_i=0 $. Then
\begin{equation}
{{{\boldsymbol \omega}}}=({\omega}^1, {\omega}^2~...~ {\omega}^k, {\omega}^{k+1})^T,
\end{equation}
where ${\omega}^1= {{\sigma}_1^{0}-{\sigma}_1^{1} \over 2}$, $ {\omega}^i = {{\sigma}_{i-1}^{0} + {\sigma}_{i-1}^{1} + {\sigma}_{i}^{0} - {\sigma}_{i}^{1} \over 2} $ for all $ i=2,...,k $ and $ {\omega}^{k+1}={ {\sigma}_{k}^{0} + {\sigma}_{k}^{1} \over 2}$, also  
\begin{equation}\label{Eq:ChallengeFeature}
\bf{{\Phi}} (\bf{{C}}) = ({\Phi}^1(\bf{{C}}),...,{\Phi}^k(\bf{{C}}),1)^T,
\end{equation}
where $ {\bf {\Phi}^j}(\bf{{C}})={ {\rm \Pi}}_{\it i=j}^{\it k}({\rm 1-2}{\it c}_i) $ for $ j=1,...,k $.
\subsubsection{Reliable Response Determination}
Suppose the $ {\boldsymbol \omega} $ is known, given a challenge $ \bf C $, the $\bf{ {\Phi}} $($ \bf C $) is determined. Then the $ t_{\rm dif} $ is calculated. Noting that $ t_{\rm dif} $ eventually comprises two important useful information: i) sgn($ t_{\rm dif} $) determines the binary response; ii) the reliability of this response. If the $ t_{\rm dif} $ is far alway from zero, then this gives such a challenge with full confidence to reproduce the response without any erroneous.

In practice, physically measuring the $ t_{\rm dif} $ is hard, if not possible. Xu {\it et al.} ~\cite{xu2016using} recently exploit machine learning techniques, specifically Support Vector Machine (SVM), to learn the $ {\boldsymbol \omega} $ using a small collection of CRPs. Once an accurate $ {\boldsymbol \omega} $ is learned through the SVM, the $ t_{\rm dif} $ is able to be accurately predicted given an unseen $ \bf C $. Then the corresponding response and its associated reliability is judiciously determined. If a challenge results in a $ t_{\rm dif} $ that is far way from zero, then its corresponding response is error-free. Xu {\it et al.}~\cite{xu2016using} demonstrate that almost 80\% randomly given challenges guarantee error-free responses across a wide range of operating conditions (temperature, voltage) as well as considering aging effects. 

However, exploiting those error-free responses, especially, in a secure manner was not considered. We take the first step towards to securely exploiting those error-free responses to construct a strong PUF
\subsection{Controlled PUF}\label{Sec:ControllPUF}
\begin{figure}
\centering
\includegraphics[trim=0 0 0 0,clip,width=0.4\textwidth]{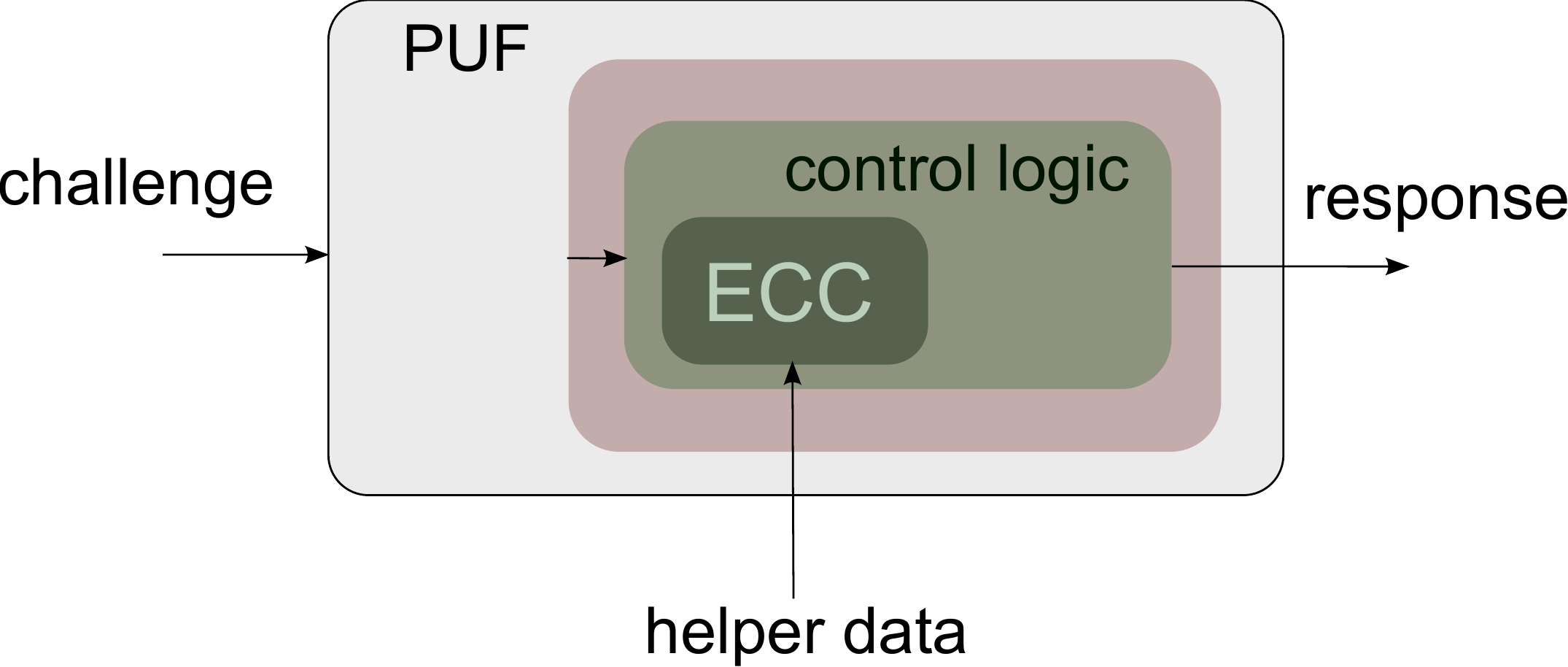}
\caption{Generalized controlled PUF construction. 
}
\label{fig:CPUF}
\end{figure}

The controlled PUF~\cite{gassend2002controlled,gassend2008controlled} proposed by Gassend {\it et al.} is a strong PUF construction. A controlled PUF is a PUF that is combined with a control logic limiting the ways in which the PUF can be evaluated. In general, without permission from a trusted entity, the controlled PUF is locked, no response will be meaningfully evaluated. When a user is authorized a CRP, more CRPs can be extracted. This is alike key management, where more session keys can be derived from a master key. In practice, the controlled PUF is built as a means that the PUF and its control logic play complementary roles. As illustrated in Fig.~\ref{fig:CPUF}, the PUF prevents invasive attacks on the control logic, at the same time, the control logic protects the PUF from protocol level attacks. For example, the APUF delay wires wrap the control logic. If invasive attacks attempt to probing the control logic, it is more likely that the PUF secret will be altered and damaged. The control logic halts adaptively evaluations on PUFs with no permission from the trusted entity.

The responses in the controlled PUF have to be post-processed, e.g., hashed. Previous works~\cite{gassend2002controlled,gassend2008controlled} usually held an assumption that the error correction code (ECC) logic and the associated helper data are default parts of a PUF. In practice, the ECC logic and storing of helper data are always expensive, especially for most low-end IoT devices. In addition, availability of the helper data is a non-trivial task in practice, especially when the key renewal is occurred. In this context, the user randomly picks up a seed challenge and queries the output---e.g., hashed responses---from the controlled PUF. Fully characterization of all possible CRP given a PUF that has exponential number of CRPs, in particular the popular employed APUF, and sequentially computing all possible helper data is infeasible. The helper data given the user randomly chosen challenges cannot be always guaranteed. Most importantly, usage of helper data poses the controlled PUF under potential threaten from modeling attacks exploiting noise side-channel information leakage~\cite{becker2014active,delvaux2015helper,becker2015pitfalls}.

The PUF-FSM is the first practical controlled PUF without using ECC along with the associated helper data and with an explicit countermeasure to reliability-based fault attacks.
\subsection{Finite State Machine (FSM)}
Finite state machine (FSM) is a popular sequential logic. In a FSM, the next state depends on both the input (transaction edge) and the current state. The FSM has been employed for IC active metering~\cite{koushanfar2001hardware,koushanfar2012provably,zhang2015puf}. In this context, the FSM combines with a unique chip identifier, usually a weak PUF, where PUF responses act as transaction edges to unlock a function such as an Intellectual Property (IP). The PUF response given a challenge, here, act as a secret key. Previous works~~\cite{koushanfar2001hardware,koushanfar2012provably,zhang2015puf} extract a constant secret or a key from the noisy PUF responses. Please note that requirements on the on-chip ECC and helper data are still existed. 

Our work employs the FSM as a control logic to realize the said controlled PUF. We release large number of challenge secret pairs. Neither ECC nor the helper data is necessary. Beyond IC active metering, our work enables authentication, key generation, key exchange and more advanced cryptographic protocols where a shared secret is required.
\section{PUF-FSM: Design and Security Analysis}\label{Sec:PUF-FSM}
\subsection{PUF-FSM Structure}
 \begin{figure}
 	\centering
 	\includegraphics[trim=0 0 0 0,clip,width=0.3\textwidth]{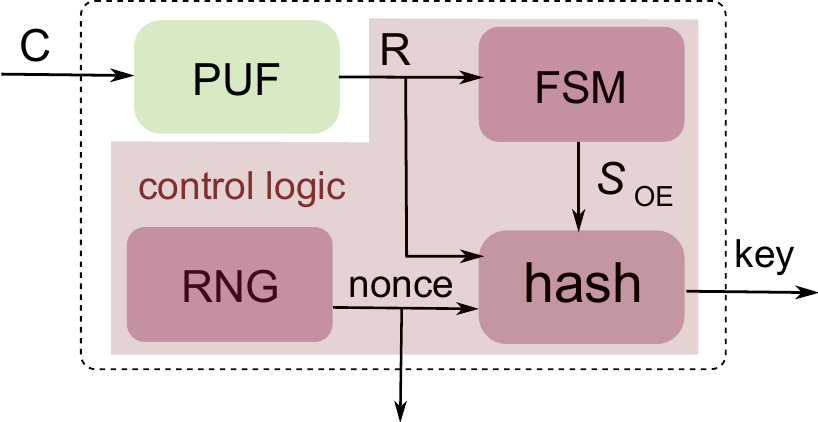}
 	\caption{General structure of the PUF-FSM. Only the correct sequential challenges produced $ \bf R $ can unlock the $ S_{\rm OE} $. If the enable signal $ S_{\rm OE} $ is disabled, the hash output is meaningless by presenting random values. Otherwise, the key is generated based on {\it part} of the response $ \bf R $, $ {\bf R}_{\rm secret} $, {\it and} the nonce, where key={\scshape hash}($ {\bf R}_{\rm secret} $, nonce).}
 	\label{fig:PUFFSM}
 \end{figure}
The PUF-FSM structure is generalized in Fig.~\ref{fig:PUFFSM}. It consists of a PUF, a FSM, a hash and a random number generator (RNG) block. Similar to priori work~\cite{yulockdown}, the direct PUF responses can only be evaluated by the trusted entity in a secure environment to build APUF statistical model(s), and the direct access is destroyed afterwards, e.g., through fusing the wire.

During deployment, a set of $ n $ sequential challenges, $ {\bf C}_{\rm set} $, is issued by the trusted entity, e.g., the server, the corresponding error-free responses $ \bf R $ with length $ n $ is produced. The $ \bf R $ is sequentially fed into the FSM controlling the transitions of the FSM states. Note that before the operation, the FSM resets to $ S_0 $. Only a series of correct $ \tt TR $---sub-response enabling the state traverse from the current state to the next state---is able to guarantee the FSM transitioning into the $ S_{\rm OE} $ that is an activation to unlock the key output. In this context, only the server who owns the statistical APUF model is capable of issuing a correct challenge set, $ {\bf C}_{\rm set} $ to unlock the $ S_{\rm OB} $ to generate a meaningful output as a key. The key is {\scshape hash}($ {\bf R}_{\rm secret} $, nonce), the $ {\bf R}_{\rm secret} $ is partial of $ \bf R $ and formation of $ {\bf R}_{\rm secret} $ will be described and clear soon. Whenever the $ S_{\rm OE} $ is disabled, the output presents random values.
 
\begin{figure}
 	\centering
 	\includegraphics[trim=0 0 0 0,clip,width=0.48\textwidth]{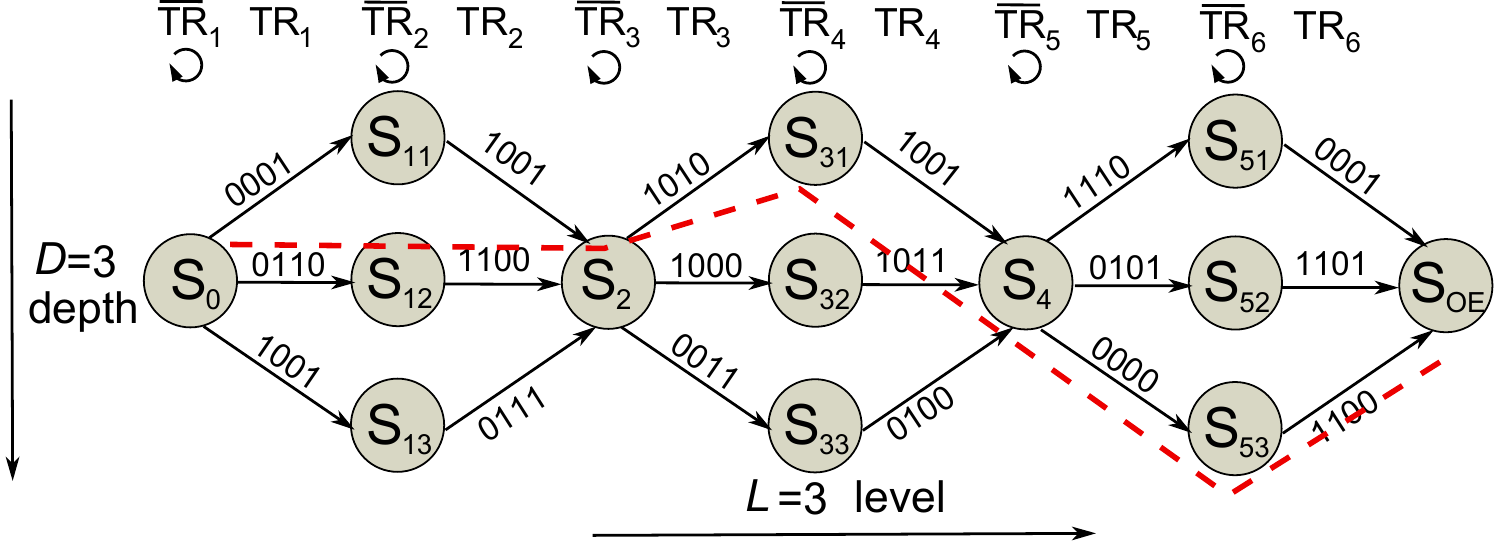}
 	\caption{FSM example with five levels ($ L=5 $) and three depths ($ D=3 $). When the transition edge $ {\tt TR}_{\rm ld} $, eg., 1100, is fed, current state $ S_{\rm (l-1)d} $, eg., $ S_{12} $, transitions into $ S_{\rm ld} $, eg., $ S_{22} $. The applied $ {\overline {\tt TR}}_{\rm ld} $ remains the FSM at its current state, marked by the returning arrow.}
 	\label{fig:FSM}
 \end{figure}

An exemplary FSM construction is depicted in Fig.~\ref{fig:FSM}. At the beginning of the PUF-FSM operation, the FSM resets to its initial state $ S_0 $. Let's assume that the $ {\tt TR}_1 $ is 0110, then $ S_0 \xrightarrow{0110} S_{\rm 12} $. Similarly if the $ {\tt TR}_1 $ is 0001, then $ S_0 \xrightarrow{0001} S_{\rm 11} $. If $ {\tt TR}_1 $ is from none of $ \{0001, 0110, 1001\} $, or in other words the $ \overline {\tt TR}_1 $ is fed, then $ S_0 \xrightarrow{\overline {\tt TR}_1} S_{\rm 0} $. Note that when the $ \overline {\tt TR} $ is fed, the FSM remains at its current state. In this case example, for even states $ S_0,S_2,S_4 $, the $ {\tt TR}_l $ having $ D $ transition edges that can lead it to any of the following $ D $ states, rest $ {\tt TR}_l $ have only one correct transition edge that leads it to the following state.

\begin{figure}
  	\centering
  	\includegraphics[trim=0 0 0 0,clip,width=0.40\textwidth]{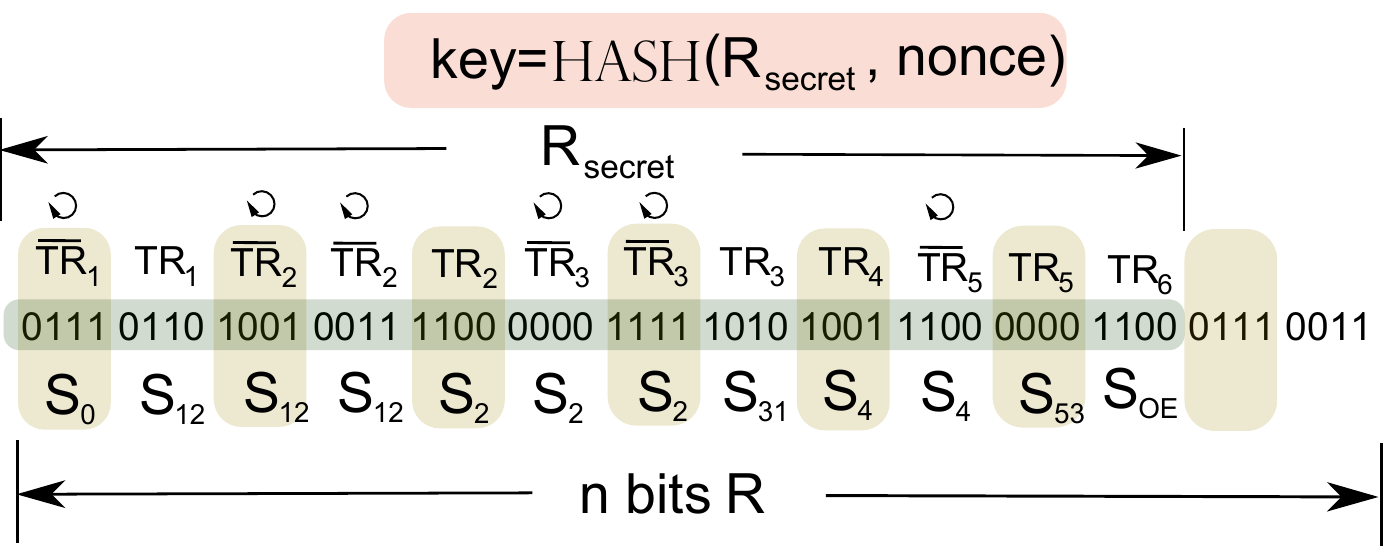}
  	\caption{ Part of $ n $-bit $ \bf R $, $ {\bf R}_{\rm secret} $ is hashed to generate the key. All rest bits after reaching the enable signal $ S_{\rm OE} $ are not contributing to the key. Note that the FSM example in Fig.~\ref{fig:FSM} is used for state traverse illustration that is marked by the dotted red line.}
  	\label{fig:Hash}
  \end{figure}

Though other FSM structures can be envisioned, the FSM in our proposal has $ L $---always an odd number---internal state layers (levels); each odd internal layer has $ D $ parallel states. A constant number, $ L+1 $, of $ {\tt TR}_l $ is a must to reach to the $ S_{\rm OE} $. Both the $ {\tt TR}_l $ and $ \overline {\tt TR}_l $ are 4-bit in this case example, therefore, the number of $ {\tt TR}_l $, and the {\it maximum} number of $ \overline {\tt TR}_l $, $ \overline n_{\rm lmax} $, in together is $ \frac{n}{4} $, where we assume that the $ n $ is always a multiplication of 4 for convenience. In practice, the $ S_{\rm OE} $ can be activated in a way by applying $ L+1 $ $ {\tt TR}_l $ and $ \overline n_{\rm l} $ $ \overline {\tt TR}_l $, noting that $ {\overline n_{\rm l}\le {\overline n_{\rm lmax}}} $. The meaningful key will be given only after all $ n $ bits in $ \bf R $ are fed into FSM---or $ n $ clock cycles past-- {\it and} the $ S_{\rm OE} $ is activated/reached. The key is a hash function of part of the $ \bf R $ that is the all sequentially fed $ L+1 $ $ {\tt TR}_l $ and $ \overline n_{\rm l} $ $ \overline {\tt TR}_l $. An example illustration of the key formation is shown in Fig.~\ref{fig:Hash}---the state traverse path is illustrated in Fig.~\ref{fig:FSM} in the dotted red line. Once the $ S_{\rm OE} $ is reached, the rest $ \overline {\tt TR}_l $ are neglected---will not be hashed to generate the key. It is worth to stress again that the rest response bits are still fed into the FSM as redundant bits to hide the length of $ {\bf R}_{\rm secret} $.
  
\subsubsection{\bf Device Nonce} The device nonce is exploited to prevent observing repeatedly evaluated responses given the same challenge~\cite{becker2015gap,becker2015pitfalls,delvaux2015helper}. The security rationale shall be clear in Section~\ref{Sec:SecurityAnalysis}. Nonce is part of the key, where the key={\scshape hash}($ {\bf R}_{\rm secret} $, nonce). It is reminded that the key will differ under each evaluation considering the freshed nonce even the same $ {\bf C}_{\rm set} $ issued by the trusted entity is repeatedly applied. The nonce is visible, the security relies on the $ {\bf R}_{\rm secret} $.

\subsubsection{\bf Design Highlights} (1) Only under a correct set of sequential challenges, $ {\bf C}_{\rm set} $, the final state $ S_{\rm OE} $ of the FSM can be reached or activated; (2) the number of $ \overline {\tt TR}_l $, $ \overline {n}_l  $, before reaching the $ S_{\rm OE} $ and the number of $ \overline {\tt TR}_l  $, ${\overline {n}_{\rm lmax}}- {\overline {n}_l} $, after reaching the $ S_{\rm OE} $ are flexible configured that is controlled and only known by the trusted entity; (3) a meaningful key is presented only when the $ S_{\rm OE} $ is activated and all $ n $ error-free response bits are fed into the FSM. If the $ S_{\rm OE} $ is disabled, a random value is presented; (4) device nonce is employed to prevent repeatedly responses' observations given the same maliciously applied challenge.
\subsection{Security Analyses}\label{Sec:SecurityAnalysis}
\subsubsection{Adversary Model}  We consider the same assumption for controlled PUFs~\cite{gassend2002controlled,gassend2008controlled} that physical attacks on the control logic is more likely to alter or even destroy the PUF itself. The adversary can eavesdrop the communication channel and arbitrarily apply challenges to the PUF-FSM input to observe the PUF-FSM output. Furthermore, the nonce is visible. The adversary attempts to obtain useful information to learn the APUF model in the PUF block. 
\subsubsection{Brute-force Attacks}
As for an adversary, the probability of discovering a meaningful key through guessing a correct $ {\bf C}_{\rm set} $ without the assistance from the trusted entity is expressed:
\begin{equation}\label{Eq:Complexity}
{\tt Probability}=(\frac{D}{2^{n_{\rm TR}}})^{\frac{L+1}{2}}\times(\frac{1}{2^{n_{\rm TR}}})^{\frac{L+1}{2}},
\end{equation}
where the $ n_{\rm TR} $ is the length of $ {\tt TR}_l $.  In the case example of Fig.~\ref{fig:FSM}, the $ n_{\rm TR} $ is four. For each even layer, the probability of guessing one correct transition edge is $ (\frac{D}{2^{n_{\rm TR}}}) $, while the probability of guessing a correct transition edge for given an old layer is $ (\frac{1}{2^{n_{\rm TR}}}) $.

The brute-force attack becomes computationally infeasible as the FSM state layer $ L $ or the $ n_{\rm TR} $ increases. In addition, even an adversary luckily guesses a correct $ {\bf C}_{\rm set} $ that unlocks the $ S_{\rm OE} $, (s)he is actually incapable of recognizing it. Output from the PUF-FSM looks random to the adversary under each evaluation without prior knowledge of a correct $ {\bf C}_{\rm set} $ attributing to the refreshed nonce.
\subsubsection{Modeling Attacks}
The plausible attacks on strong PUFs are modeling attacks. Numerous works ~\cite{ruhrmair2013puf,ruhrmair2010modeling,majzoobi2008testing,becker2015gap,becker2015pitfalls} have shown the vulnerability of the strong PUFs to modeling attacks. Those deemed but later breakable strong PUFs include XOR-APUF, Feedforward APUF, Lightweight Secure PUF and even Slender PUF. 

In PUF-FSM, arbitrarily CRP collection is disabled by any party except the trusted entity during the secure enrollment phase. After the enrollment phase, the response is never directly exposed unless hashing and its usage is further controlled by the FSM. The control logic as shown in Fig.~\ref{fig:PUFFSM} first protects the underlying APUF(s) from modeling attacks such as LR and SVM where knowledge of responses is necessary~\cite{ruhrmair2013puf,ruhrmair2010modeling}. 

As to perform recent revealed modeling attacks exploiting the helper data information~\cite{delvaux2015helper,becker2015pitfalls}, in other words, the unreliability information of a given CRP, knowledge of which challenge is unreliable is a premise. Unlike traditional modeling attacks, e.g, LR, reliability-based fault CMA-ES attacks~\cite{becker2015pitfalls} do not require the knowledge of the response value for a given challenge. Such a powerful CMA-ES attack even threatens the security of a controlled PUF that employs the helper data. In our PUF-FSM, there is no helper data involved. Thus, exploitation of information leakage from the helper data to perform reliability-based attacks is excluded.

Now without using the device nonce, we examine the means of finding unreliable challenges by observing the PUF-FSM output rather than gaining information from the helper data.
By applying arbitrarily challenges to the PUF-FSM and without priori knowledge of a correct $ {\bf C}_{\rm set} $, there is no information that can be observed and used by the adversary to discover the unreliable challenges. This lies on the fact that the output of the PUF-FSM is random or meaningless, if the enable signal $ S_{\rm OE} $ is locked/disabled. The complexity of unlocking the $ S_{\rm OE} $ without the participation of the trusted entity is same to the brute-force attacks given in (\ref{Eq:Complexity}). 

We note that there still exists a potential way to determine an unreliable challenge through exhaustive search under the assumption that a priori $ {\bf C}_{\rm set} $ has been eavesdropped and now the adversary is holding the physical PUF-FSM. The adversary chooses an unused challenge $ {\bf C}_x $ to replace one challenge $ {\bf C}_i $ in the eavesdropped $ {\bf C}_{\rm set} $ to observe the output of the PUF-FSM. If $ {\bf C}_x $ is an unreliable challenge and its response contributes to the $ {\tt TR} $. Then under repeatedly evaluations, the adversary can determine such an unreliable challenge when the key and random output are iteratively exhibited. If $ {\bf C}_x $ is unreliable and its response contributes to the $ \overline {\tt TR}_l $. Then under repeatedly evaluations, an unreliable challenge is determined when two differing keys are iteratively exhibited. Through continuous exhaustive searching, other unreliable challenges can be determined as well to perform reliability-based attacks.

By employing the device nonce alike~\cite{yulockdown}, no matter the $ {\bf C}_x $ is unreliable or not, due to the nonce being refreshed each evaluation, observing the same key by repeatedly applying the same challenge is infeasible. Thus, discovery of the unreliable challenge is disabled. The reliability-based attack~\cite{becker2015gap,becker2015pitfalls} is, as a result, prevented.
\section{Applications}\label{Sec:Discussion}
\subsection{Mutual Authentication}
The PUF-FSM achieves mutual rather than common unidirectional authentication. Recall that only a trusted entity has the capability of issuing a correct challenge sequence to activate the $ S_{\rm OE} $. As a consequence, only the PUF-FSM device and the trusted entity know the $ S_{\rm secret} $. 

When the PUF-FSM is transfered to the user. The trusted entity issues a $ {\bf C}_{\rm set} $ and sends them to the user may through insecure communication channels. The user presents the $ {\bf C}_{\rm set} $ to the PUF-FSM and sends both the nonce and the PUF-FSM output (key) back to the trusted entity. The trusted entity {\it computes} a key, {\scshape hash}($ {\bf R}_{\rm secret} $, nonce), and compares it with the key received. If they are same, the user holding the PUF-FSM is authenticated. Once the user is authenticated, the user applies the same $ {\bf C}_{\rm set} $ again to the PUF-FSM to obtain a refreshed output (key). The user asks the refreshed key computed by the trusted entity after sending out the nonce. The trusted entity is authenticated by the user only if the received computed key is same to the key produced by the PUF-FSM.

\subsection{Key Exchange}
Following the foregoing mutual authentication, we consider the key exchange scenario between the user and the trusted entity.  The user applies the same $ {\bf C}_{\rm set} $ and sends the nonce to the trusted entity. But there is no key (shared key) sending between two parties. Now only the user who holds the PUF-FSM and the trusted entity know the shared key. The user obtains it from the PUF-FSM, while the server {\it computes} it by hashing the $ {\bf R}_{\rm secret} $ and the nonce.
\subsection{Controlled PUF}
Served as a controlled PUF, intermediate benefits of the PUF-FSM are the exclusion of the on-chip ECC logic and the usage of helper data, which finally release the constraints on a practical realization of the controlled PUF. In addition, no ECC and helper data eliminates potential security concerns on previous controlled PUF designs from modeling attacks, where the helper data leaks information~\cite{becker2015pitfalls,becker2014active,delvaux2015helper}.

\subsubsection{Key Obtain} As an  intentional design purpose, the controlled PUF restricts the means in which a PUF can be evaluated. Who holds the PUF-FSM is unable to evaluate it to obtain a $ \langle {\bf C}_{\rm set}, {\bf R}_{\rm secret}\rangle $---$ \langle, \rangle $ means a tuple---without permission from the trusted entity. To acquire a  $ \langle {\bf C}_{\rm set}, {\bf R}_{\rm secret}\rangle $, first, the mutual authentication is performed to establish trustiness between the trusted entity and the user who needs to hold the physical PUF-FSM. Then the trusted entity issues a fresh set of challenges to the user who is now authorized with a $ \langle {\bf C}_{\rm set}, {\bf R}_{\rm secret}\rangle $.

\subsubsection{Key Renewal} Once the user is authorized with a $ \langle {\bf C}, {\bf R}_{\rm secret}\rangle $, (s)he is able to renew arbitrary keys from the PUF-FSM. The $ {\bf R}_{\rm secret} $ can be treated as a master secret, where all other sub-keys, {\scshape hash}($ {\bf R}_{\rm secret} $, nonce), are available. Given a known nonce, the user and the trusted entity can retrieve sub-keys or sub-session keys without issuing a new challenge set. A shared key between two parties indeed enable a wide variety of standard cryptographic protocols to be implemented~\cite{gassend2008controlled}.
\section{Conclusion}\label{Sec:Conclusion}
We have presented a practical controlled strong PUF, PUF-FSM, by (1) exploiting error-free responses in absence of an APUF and (2) controlling the means of evaluating the PUF by using a control logic. The PUF-FSM requires neither on-chip ECC nor helper data that were usually must when extracting a key. As a controlled PUF, it holds the promise of a cost-effective way to increase resistance to various attacks, especially invasive attacks, for IoT devices. Security analyses demonstrate that the PUF-FSM is resilient to modeling attacks.


\end{document}